# Dynamic Address Allocation Algorithm for Mobile Ad hoc Networks

Akshay Khatri[1], Sankalp Kolhe[1], Dr. Nupur Giri[1]

[1]Vivekanand Education Society's Institute of Technology, Mumbai, India

A Mobile Ad hoc network (MANET) consists of nodes which use multi-hop communication to establish connection between nodes. Traditional infrastructure based systems use a centralized architecture for address allocation. However, this is not possible in Ad hoc networks due to their dynamic structure. Many schemes have been proposed to solve this problem, but most of them use network-wide broadcasts to ensure the availability of a new address. This becomes extremely difficult as network size grows. In this paper, we propose an address allocation algorithm which avoids network-wide broadcasts to allocate address to a new node. Moreover, the algorithm allocates addresses dynamically such that the network maintains an 'IP resembles topology' state. In such a state, routing becomes easier and the overall overhead in communication is reduced. This algorithm is particularly useful for routing protocols which use topology information to route messages in the network. Our solution is designed with scalability in mind such that the cost of address assignment to a new node is independent of the number of nodes in the network.

## I. INTRODUCTION

A variety of addressing schemes have been proposed to efficiently allocate addresses to nodes in Mobile Adhoc Networks (MANETs). Following are the desirable properties for addressing schemes in a MANET:

1. *Desired Properties of an Addressing Scheme:*

- *Unique address allocation*

A unique address should be allocated to all nodes joining the network. There are 2 ways to ensure this: Duplicate Address Detection (DAD) i.e. to make sure an address is available before assigning it to a node, or using disjoint address tables to make sure that there is no need for a DAD.

- *Low overhead*

The communication overhead for address assignment should be low. This means that network-wide broadcasts for address allocation should be avoided.

- *Network partitions*

The addressing scheme should allocate addresses in such a way that network partitions can be handled easily.

- *Network joins*

Network joins are generally associated with a change of address for a large number of nodes. The addressing algorithm should work well in this scenario or else it may cause a congestion in the network.

- *Zero configuration*

The algorithm should configure itself automatically with minimum human involvement. This means that address allocation and claiming free addresses should be done automatically and dynamically.

- *Robust towards moving nodes*

In MANETs, node move frequently which means that the path from node A to node B will change continuously. The addressing scheme should make this process of finding new paths easier.

In this paper, we propose an Address allocation algorithm that attempts to perform the tasks of address allocation and management in a scalable way by localizing the communication between nodes for these tasks. We do this by forming local clusters in the network and assigning a pool of addresses to this cluster. From here, it is the cluster's responsibility to maintain the address table for this pool of addresses.

## II. RELATED WORK

In Prophet Addressing [1], all nodes in the network know the IP addresses that are going to be assigned to nodes joining the network in advance. The list of IP addresses that are going to allocated are generated by a function f(n). The initial state is chosen at random and is known as seed. f(n) depends on the current IP address and seed. The seed value is updated with each node joining the network. This algorithm works on network partitions and merges. The algorithm however fails to assign collision-free address even if the address range selected is relatively large. Duplicate Address Detection(DAD) is an approach which doesn't specify the initial address assignment mechanism but instead works on verifying uniqueness of a randomly assigned address. Strong DAD is a DAD algorithm which uses flooding the network to detect an address conflict. The overhead generated due to flooding makes it unfavorable for large MANET. Weak DAD [2] improves upon Strong DAD approach by making IP address is not the only identification factor for nodes in a network. Along with IP address each node is also assigned a key and the pair of <IP address, key> is used to identify nodes. Thus even if nodes with same IP addresses exist in a network their keys would be different making them unique. A common practice is to assign the nodes MAC address as the key. However, MAC address of nodes can be same thereby eliminating the uniqueness of keys. Weak DAD handles network merger effectively and prefers the use of proactive routing algorithms which makes the network susceptible to failures and adds overhead of data maintenance. Passive DAD [3] is another DAD approach in which link state routing protocol traffic is analyzed. The nodes in the network notify other nodes about their neighbors using link state packets. This approach has broadcast overhead and the limitation of using only link-state routing. Dynamic Host Control Protocol(DHCP) uses a DHCP server to assign IP addresses to nodes. The DHCP server maintains a pool of available IP addresses. DHCP can also be used as an auto-configuration protocol. In MANETs Distributed DHCP(D2HCP) is used, in which all nodes in the network are mapped and this information is available to every node in the network. Thus a new node can be assigned an available IP address by any node in the network. However, conflict-detection algorithms need to be implemented along with D2HCP as two or more nodes may be assigned the same IP address by different nodes if they join the network at almost the same time. D2HCP eliminates the need for static configuration or centralized server however adds the overhead of maintaining and synchronizing multiple tables. Dynamic Address Configuration Protocol(DACP) [4] elects a centralized Address Authority (AA) which assigns temporary IP address to nodes joining the network and uses conflict-detection protocols to determine its uniqueness. A conflict-free temporary IP address is made permanent for a node. In DACP due to a regular conflict-detection there is a high flooding. A centralized AA also has an overhead. Optimized DACP (ODACP) [8] is an iteration of DACP and removes the overhead caused by DACP by eliminating the need for conflict-detection. However, ODACP has scalability issues owing to its centralized approach to address assignment by the elected node. MANETConf [4] is an approach to address assignment in which new node joining the network requests neighbor node in the network with its desirable IP address. The neighbor node broadcasts the IP address in the network to ensure its uniqueness. MANETConf overcomes the constraint of simultaneous requests by different nodes by using pending allocation tables. MANETConf handles network partitions efficiently. The major drawback of this protocol is that nodes have to maintain tables for the entire network. There are also frequent and timely broadcasts in this protocol which add to the existing overhead. Distributed Protocol for Dynamic Address [7] is a dynamic address assignment scheme makes use of the concept of buddy system used for memory management. In this protocol nodes in the network have unused IP address blocks known as free_ip. A new node joining the network is assigned the second half of the free_ip block by a node in the network. Thus this protocol does not need centralized servers. In this protocol latency and communication overhead is dependent on the address block size. In absence of large address block this protocol faces scalability issues. The idea of Dynamic addressing was introduced by Eriksson et.al in DART [5]. They develop the routing algorithm and an addressing scheme based on Dynamic addressing. For the addressing part, they take the new address from the neighbor's largest unoccupied address block. This node starts functioning in the network and address validation is done through a routine that ensures that the prefix subgraph constraint is not violated. In case this constraint is violated in the network, the node takes a new address. Our work on the other hand performs the addressing dynamically and in a distributed way. We introduce the idea of cluster heads which maintain the address tables for a small subset of nodes in the network. These nodes collectively perform the address management in the network.

III. BASIC IDEA

In traditional Adhoc networks, when a node joins a network, it is assigned a unique IP address which is retained throughout its lifetime in the network. The IP address is the identifier for the node in the network. Thus, traditional systems follow an 'address equals identity' model which is inherited from wired systems [5]. But, a node's IP address can provide information about its location in the network, rather than just being a unique identifier. This can be achieved via dynamic addressing. Dynamic addressing does not follow the model of 'address equals identity'. It uses some other parameter as the unique identifier, say MAC address. It emphasizes on an 'IP address resembles topology' model which means that the arrangement of nodes in the network is represented by the IP addresses of the nodes in the network. Dynamic addressing, as the name suggests, allocates and deallocates IP addresses dynamically in such a way that nodes which are close to each other have a similar address. Moreover, as nodes move in the network, they change their addresses to be coherent with their new location in the network. This makes routing easier since the IP address can help the routing protocol to find the path to destination node easily.

Distributed addressing is a stateful approach which uses distributed maintenance of multiple disjoint allocation tables for addressing wherein the responsibility of assigning addresses and maintaining the address table does not rely on a single node [6]. Relying on a single node for these tasks is not good for a number of reasons. First, it creates high load on the agent which is doing the task of allocation. Second, the network is highly dependent on a single node i.e. it becomes centralized. Distributed addressing avoids these problems by distributing this task of address assignment among a number of nodes.

We have taken the ideas of Distributed and Dynamic addressing and used them together to propose a solution that is decentralized, has low communication overhead and makes routing of messages easier in the network. We use the hierarchical nature of IP addresses to generate a hierarchy in the network wherein each node maintains the details of the nodes in the level just below its own level. For the sake of simplicity, let us consider an IPv4 network with a range of addresses from 10.1.0.0 to 10.1.255.255. Our algorithm is equally applicable to IPv6, or for a network with a different subnet mask. We define levels in the network for the purpose of defining the hierarchy in the network and the level to which a node belongs can be identified from it IP address.

We use each of the four 8-bit parts of the IP address to define levels in the network. We define that a node is at level k if it has k continuous zeros in its IP address (written in a dotted decimal form) when traversed from right, without considering zeros in the network prefix. Therefore, 10.1.0.0 is at level 2 whereas 10.1.1.0 is at level 1 and 10.1.1.1 is at level 0, where the network prefix is 10.1. A node at level k acts as the *"cluster head"* for up to 255 nodes at level k-1 and maintains an address table to store the allocated IP addresses and their respective MAC addresses. A node at level 0 does not have to maintain a table to store this information since there are no levels below 0. Nodes which are on the same level and have the same cluster head belong to the same cluster.

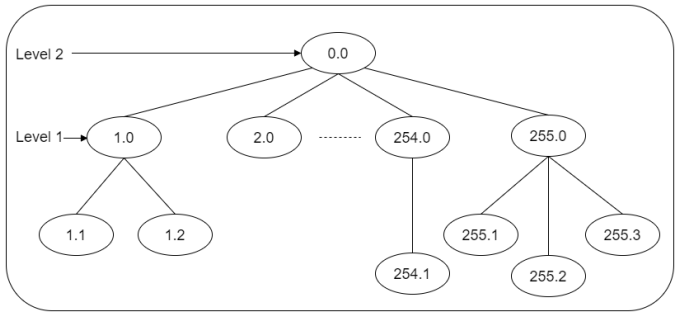

Fig. 1. A snapshot of the proposed network

For example, in Fig. 1, the node with the address 10.1.0.0 is the cluster head of the nodes on level 1 i.e. nodes with addresses from 10.1.1.0 to 10.1.255.0. In other words, nodes with addresses from 10.1.1.0 to 10.1.255.0 (level 1) form a cluster and their cluster head is 10.1.0.0 (level 2). Similarly, the node with the address 10.1.1.0 is on level 1 and is the cluster head of 255 nodes at level 0 i.e. nodes in the range 10.1.1.1 to 10.1.1.255. Given this address space we can accommodate 65536 nodes in the network, but it is possible to extend beyond this number by changing the subnet mask.

### A. Connectivity Constraint:

We define a connectivity constraint for nodes at level 0. At any point of time, each node at level 0 should be directly connected to at least one node from the same cluster. If the node is not directly connected to at least one node from the same cluster for a certain threshold period of time, it will look for nodes of another cluster which are directly accessible and request the node for an address change and it will become a member of that cluster. The constraint is defined at level 0 because only members of the 0 level cluster need to be in close proximity of each other. For higher levels, it is not feasible for the nodes of the same cluster to be directly accessible to each other.

Consider this analogy: suppose each of the node at level 0 represents a house in a town and its cluster head is the mayor. While each of the houses (level 0 nodes) in a town would be in close proximity to at least one other house, all the towns may not be on close proximity to each other. In other words, it is not possible for all the mayors (level 1 cluster heads) to be in direct contact to each other.

### IV. NETWORK INITIALIZATION

The network is initialized by a single node. The node sends out broadcasts to check if a network exists that it can join. If it does not receive a response for a certain period of time, it concludes that a network does not exist and initializes one. The node selects a network prefix and a subnet mask which decides an address range, which by default is 0.0 to 255.255. The initializing node takes the first address from the address pool which in this case is 10.1.0.0. The initializing node also selects a unique network id for the network, which is derived from its MAC address. We call this unique identifier '*NetID*'. The network id remains the same throughout the lifetime of the network. The node which initializes the network will be at the highest level in the network. We call this node as the '*supreme*' node. In this case, the supreme node is at level 2.

```
Pseudocode for Network Initialization:
def network_init():
  while(!Timeout):
    self.broadcast(join_network_request)
      if reply == 1:
        self. get_address()
        return()
  NetID = 10.1
  // Optional: choose an address range
  self.address = NetID + "0.0"
  self.address_table = []
```

### V. ALGORITHM AND WORKING OF THE PROTOCOL

Each cluster head creates and maintains an address table which lists all the addresses in its range and whether they are taken or free. The address table also includes the MAC address of the node and the most recent timestamp when the node has reported to the cluster head. The address table is copied and regularly updated at other nodes as well so that the address table is not lost when the cluster head fails. We call these nodes subordinate head nodes. These nodes act as a backup to the address table information which is being stored by the cluster head. These subordinate heads are the nodes which become the cluster head if the cluster head leaves the network. Subordinate nodes are generally directly accessible from the cluster head.

All nodes in the network send periodic updates to their cluster heads, so that the cluster heads can keep track of all the active nodes in the cluster and free up the addresses of nodes which are no longer active in the network. This makes sure that every cluster head's address table is up-to-date at any point of time. The advantage of such a design is that all the protocol messages are local broadcasts and are not sent outside the cluster. This reduces the communication overhead in address management and makes the solution scalable.

The main objective of the protocol is to maintain the network in an 'IP resembles topology' state. This means that all the nodes have to keep a check on their surrounding nodes so that such a state is maintained. Nodes at different levels have different ways of handling this.

The connectivity constraint is enforced at level 0. Each level 0 node periodically scans its surroundings to ensure that it is directly connected to at least one node from its cluster. If a node finds out that it is no longer connected to its cluster, but it is connected to a node from another cluster, it makes a change address request. The details of this process are explained in later sections.

At higher levels, each node tries to make sure that it is directly connected to at least one node from the cluster whose address details it maintains. If the node has lost direct contact to that cluster, it moves to another cluster by changing address and the subordinate head is informed so that it can become the cluster head by changing its address.

## VI. Assignment of Address to a New Node

When a node wants to join an existing network, it does so by finding an allocator. An allocator is a node in the network which is directly accessible to the requesting node. It will help the new node to obtain an IP address and get registered in the network. If a node is directly connected to multiple nodes, it will choose the node at the lowest level as its allocator.

If allocator is a cluster head at level k, it will not query its cluster head, but it will itself do the address allocation if it has a free address. The requestor will be allocated an address from the allocator's pool and hence it will have a k-1 level address. If the allocator does not have a free address, it will query its cluster head for a free address, as specified earlier.

The most frequent case will be when the allocator will be a level 0 node. In this case, the allocator will accept the request and query its cluster head for a free address. Two cases are possible:

### A. Cluster head has a free address in its address table

If the cluster head has an available address in its address table, it will notify the allocator about the same and send a message to the allocator including the available address. The allocator will forward this message to the requestor so that it may take that address and join the network. Once this is done, the requestor will inform its MAC address to the allocator so that the address table can be modified. The allocator will send an acknowledgement to its cluster head which will include the MAC address of the requestor. On receiving the acknowledgement, the cluster head will modify its address table to accommodate the new node and send a 'process complete' message to the allocator. On receiving the final acknowledgement, the allocator node will send a message to the requestor and the requestor can now send and receive messages in the network.

### B. Cluster head does not have a free address in its address table

If the cluster head does not have a free address, it will query its cluster head which will be a higher level node. This will continue until a free address is found. In this case, the new node will not receive a level 0 address, it will be a higher level address and the node will be a cluster head. It will initialize an empty address table to store details of new nodes which may join the network.

```
Pseudocode for Address Assignment:
Requestor:
def get_address():
  nearby_nodes = self.find_nearby_nodes()
  allocator = nearby_nodes[0]
  for node in nearby_nodes:
    if node.level < allocator.level:
      allocator = node
allocator.request_address()

Allocator:
def request_address():
  //Allocator is a cluster head
  if self.level > 0:
  //Case 1: Allocator has free ip
      if len(self.address_table) < 255:
        self.assign(requestor, ip)
  else:
    //Case 2: Allocator has no free ip or it
    // is not a cluster head
    ip = self.request(self.cluster_head)
    self.assign(requestor, ip)
```

## VII. Node Departure

The self-healing action taken after the departure of a node depends on the level of the leaving node, whether it is a level zero node or not. Moreover, nodes may leave the network in two ways, abruptly or through an exit mechanism. Abrupt departure of a node generally takes due to hardware failure. When a level zero node exits the network, the network has to reclaim the newly freed address. When a higher level node leaves the network, a new cluster head has to be chosen for the cluster to which it belonged. This new cluster head has to make sure that its address table is up-to-date. The details of this process are discussed in the next section.

First, when a level 0 node leaves the network via the exit mechanism, it notifies one of the directly reachable nodes in its cluster. We call this node 'deallocator'. The deallocator node informs the cluster head about the node's departure. The cluster head modifies the address table makes the address available.

Each level 1 cluster head receives periodic updates from level 0 nodes in its cluster. When a level 0 node leaves the network abruptly, cluster head will not receive any more updates from the departed node. After waiting for a threshold period of time, the cluster head will send a message intended for the departed node. When the cluster head does not receive a reply to the message, it concludes that the node has left and modifies the address table by deleting the MAC address corresponding to that IP, thereby making the address available for other nodes.

Second, when a higher level node leaves the network via the exit mechanism, it informs its subordinate nodes about its decision. It makes sure that the subordinate's address table is synced with its own address table. When the cluster head finally leaves the network, the subordinate node changes its IP address to become the new cluster head.

If the cluster head leaves the network abruptly, the network does not realize this until some node tries to contact the cluster head. If a node cannot connect to the cluster head, it informs the subordinate node about the cluster head's departure. The subordinate node checks to make sure that the cluster head has left the network. If the head has really left the network, the subordinate node changes its IP address and becomes the cluster head.

```
Pseudocode for Node Departure:
Exiting Node:
if self.level == 0:
    deallocator = self.nearby()
    self.notify(deallocator)
else:
    self.address_table.sync(subordinate_node)

Deallocator:
reply = deallocator.notify(exit_node,
cluster_head)
if reply == 1:
    deallocator.confirm(exit_node)

Cluster Head:
self.address_table.delete(exit_node)
self.confirm(deallocator, exit_node)
```

## VIII. IP Change due to Mobility of Nodes

The protocol ensures that the network is an 'IP resembles topology' state. However, nodes in an Adhoc network have high mobility. So, when a node is no longer surrounded by nodes with similar IP addresses, it changes its address. The condition on when a node should change its IP address depends on whether it is level 0 node or a higher level node.

As described earlier, every level 0 node regularly checks if it is directly connected to at least one other node from its cluster. If the node is not connected to its cluster, it will find a node of another cluster which is directly accessible and request a new address from that node. When the cluster head does not receive an *'alive'* message from the node for a threshold period of time, it will find out that the node has left the cluster and reclaim its address.

When a higher level node moves and is no longer connected to at least one node of the cluster whose address table it was maintaining, it will find another cluster which is accessible. After joining the new cluster, it will send a message to its subordinate node asking it to become the new cluster head.

```
Pseudocode for IP address change:
while(True):
  nearby_nodes = self.find_nearby_nodes()
  for node in nearby_nodes:
    if node.cluster == self.cluster:
      wait(threshold);
      break()
  self.address = self.get_address();
```

## IX. Node Lookup

Node lookup is the process of finding the current address of a node in the network. This step is required exclusively in dynamic addressing because the network does not follow a 'address equals identity' model and every node has a variable address.

Given a particular id of a node, the lookup is done in a recursive manner. The process is done via a *FindAddress* request. First, the node (at level k) queries its cluster head (at level k+1) to check if the node belongs to its cluster. If the node does not belong to the same cluster, the cluster head will forward the *FindAddress* request to the members in its own cluster (at level k+1). If the node is still not found, the FindAddress request is forwarded to a higher level cluster head (at level k+2). This process is repeated until the address is found or all the cluster heads have been queried. Therefore, node lookup can be done with transfer of 255*(k+1) messages in the worst case.

When a cluster head receives a *FindAddress* request from a node that is on a level lower than its own, it knows that the request has already been through on all cluster heads on lower levels. So, it forwards the request only to nodes in its own cluster i.e. on the same level. On the other hand, when a cluster head receives a *FindAddress* request from a node on the same level or higher level, it searches its own address table and then forwards the request to the nodes of cluster it maintains the details of, i.e. cluster heads on the lower level.

```
Pseudocode for Node Lookup:
def FindAddress(id):
  if self.level == 0:
    add = self.cluster_head.FindAddress(id)
  else:
    if id in self.address_table:
      add = self.address_table[id]
    else:
      heads = self.cluster_head.get_all_nodes()
      for node in heads:
        if id in node.address_table:
          add = node.address_table[id]
          return add
      add = self.cluster_head.FindAddress(id)
  return add
```

For example, if a level 0 node with an address 10.1.23.4 wants to send a message to the node with the identifier ID1, it will initiate a *FindAddress* request. First, it will query its cluster head (10.1.23.0) to check if the node belongs to its own cluster. If yes, the cluster head will send the current address of the node and the communication begins. If the node does not belong to the same cluster, the cluster head will forward the *FindAddress* request to the nodes at its level (level 1) i.e. nodes with addresses from 10.1.1.0 to 10.1.255.0. All of these nodes will check their address tables for the identifier ID1. If any of the nodes finds an entry for ID1, it will send the node's current address to the cluster head of the requesting node (10.1.23.0), which will in turn forward it to the requesting node. If this request fails, then the request is forwarded to a node at a higher level (level 2) i.e. 10.1.0.0. But, since there aren't any more nodes at level 2, it means that all the cluster heads have been queried and the node is not available in the network and the *FindAddress* request is terminated with a fail message.

## X. NETWORK PARTITIONING

Network partitioning can be handled easily with this protocol. This is because of distributed maintenance of disjoint address tables. When the network partitions, each cluster will be either a part of the new network or the old network. The supreme node which takes the first address of the pool (10.1.0.0 in this case) will join either one of these 2 networks. The other network will use the self-healing methods to automatically select a new supreme node. This new supreme node will select a new network prefix and a Network identifier - *NetID* for the network. It will broadcast this new network prefix and *NetID* throughout the new network and all the nodes will change their IP addresses accordingly.

## XI. NETWORK MERGING

The ease at which two networks can be merged depends on the number of nodes at level k-1, where k is the level of the supreme node. Let the number of nodes at level k-1 be n1 and n2 for the two networks with Network identifiers NetID1 and NetID2 respectively.

Network merging can be done in one of the following two ways:

• If the sum of number of nodes at level k-1 in the two networks is less than 255, the networks can be simply merged directly. The new supreme node can be chosen by comparing the values of n1 and n2. If n1 is greater than n2, supreme node of Network1 will be the supreme node of the newly formed network and vice versa. The other supreme node will transfer its address table to the new supreme nodes. Nodes in the other network can simply offset the k-1 level address by n1. After transferring its address table to the newly elected supreme node, the other supreme node will simply join one of the clusters in the network through a network join request. For example, consider two networks N1 and N2 with network prefixes 10.12 and 10.23 respectively. Suppose N1 has 50 nodes at level k-1 and N2 has 25 nodes at level k-1 i.e. level 1. For N1, the cluster heads at level 1 take up addresses from 10.12.1.0 to 10.12.50.0. Similarly, cluster heads at level 1 in the network take up the addresses from 10.23.1.0 to 10.23.25.0. Since N1 has a higher number of nodes at level k-1 (level 1), the supreme node of N1 is chosen as the supreme node of the new network. All the nodes in N2 change their addresses by adding 50 to the third part of their IP addresses i.e. 10.23.1.0 becomes 10.23.51.0, 10.23.23.12 becomes 10.23.73.12. Finally, all the nodes in the network N2 change their network prefix to match the network prefix of the network N1.

• If the sum of number of nodes at level k-1 in the two networks is more than 255, direct merging is not possible. There are two ways to do the network merge:

### 1) Address Space shrinking:

Shrink the number of cluster heads at level k-1 so that the total number of clusters is less than 255. This can be done by removing redundant cluster heads. Two cluster heads are redundant if the total number of nodes in both the clusters is less than 255 and these clusters are directly connected. In such a situation, these clusters can be merged thereby freeing up an address at level k-1. If the total number of nodes at level k-1 can be reduced to less than 255 through address shrinking, the two networks can then be merged as mentioned earlier.

### 2) Increasing k:

Another way to perform the network merge is to increase the value of k, if possible. If the address space of the networks allows values of k higher than the current value, the merging can be done simply by increasing k and making the supreme nodes for the networks N1 and N2 as k-1 level cluster heads. In the above example, networks N1 and N2's address space could support higher value of k i.e. 3. In such a case, it is easy to just increase k to 3 and b choosing a new supreme at level 3 with the address 10.0.0.0. Then nodes in networks N1 and N2 can then simply change the first and second parts of their IP addresses to 10.1 and 10.2 respectively.

## XII. CONCLUSION AND FUTURE WORK

We propose an efficient addressing scheme for Adhoc networks which is dynamic and distributed in nature. It can scale well up to a large number of nodes. This is because the address allocation and management overhead is localized and limited to within the cluster itself. We use the hierarchy in IP addresses so that the algorithm is simple to understand and implement. Such distributed maintenance of the address table means that a new node can obtain a new IP address quickly without the need of any duplicate address detection schemes. Moreover, a node's IP address also provides information about its location in the network which makes routing easier. Our scheme also makes Network partition and merging easy to handle. Adhoc and Mesh networks are the next generation of communication methods which are being held back by their scalability issues. Dynamic addressing can alleviate this problem and our work is a small step in this direction.

REFERENCES

[1] Zhou, H.; Ni, L.M.; Mutka, M.W., "Prophet address allocation for large scale MANETs," in *INFOCOM 2003. Twenty-Second Annual Joint Conference of the IEEE Computer and Communications. IEEE Societies*, vol.2, no., pp.1304-1311 vol.2, March 30 2003-April 3 2003

[2] Nitin H. Vaidya. 2002. Weak duplicate address detection in mobile ad hoc networks. In*Proceedings of the 3rd ACM international symposium on Mobile ad hoc networking & computing* (MobiHoc '02). ACM, New York, NY, USA, 206-216.

[3] Weniger, Kilian. "Passive duplicate address detection in mobile ad hoc networks." *Wireless Communications and Networking, 2003. WCNC 2003. 2003 IEEE*. Vol. 3. IEEE, 2003.

[4] Sun, Yuan, and Elizabeth M. Belding-Royer. "A study of dynamic addressing techniques in mobile ad hoc networks." *Wireless Communications and Mobile Computing* 4.3 (2004): 315-329.

[5] Eriksson, Jakob, Michalis Faloutsos, and Srikanth V. Krishnamurthy. "DART: Dynamic address routing for scalable ad hoc and mesh networks."*IEEE/ACM Transactions on Networking (TON)* 15.1 (2007): 119-132.

[6] Haro, F. "IP address assignment schemes for mobile ad hoc networks. "Report for the University of Catalunya (2006).

[7] Thoppian, Mansi Ramakrishnan, and Ravi Prakash. "A distributed protocol for dynamic address assignment in mobile ad hoc networks." Mobile Computing, IEEE Transactions on 5.1 (2006): 4-19.

[8] Y. Sun, E.M. Belding-Royer, A study of dynamic addressing techniques in Mobile Ad Hoc Networks, Wireless Communications and Mobile Computing (April) (2004).